# Missile Acceleration Controller Design using PI and Time-Delay Adaptive Feedback Linearization Methodology

Chang-Hun Lee[1], Jin-Ik Lee[2], Byung-Eul Jun[3]
*Agency for Defense Development (ADD), Daejeon, 305-600, Korea*

Min-Guk Seo[4], Min-Jea Tahk[5]
*Korea Advanced Institute of Science and Technology (KAIST), Daejeon, 305-701, Korea*

A straight forward application of feedback linearization to the missile autopilot design for acceleration control may be limited due to the nonminimum characteristics and the model uncertainties. As a remedy, this paper presents a cascade structure of an acceleration controller based on approximate feedback linearization methodology with a time-delay adaptation scheme. The inner loop controller is constructed by applying feedback linearization to the approximate system which is a minimum phase system and provides the desired acceleration signal caused by the angle-of-attack. This controller is augmented by the time-delay adaptive law and the outer loop PI (proportional-integral) controller in order to adaptively compensate for feedback linearization error because of model uncertainty and in order to track the desired acceleration signal. The performance of the proposed method is examined through numerical simulations. Moreover, the proposed controller is tested by using an intercept scenario in 6DOF nonlinear simulations.

---

[1] Senior Researcher, Agency for Defense Development (ADD), Yuseong, P.O.Box 35, Daejeon, 305-600 Korea/chlee@fdcl.kaist.ac.kr.
[2] Principal Researcher, Agency for Defense Development (ADD), Yuseong, P.O.Box 35, Daejeon, 305-600, Korea/jinjjangu@gmail.com.
[3] Principal Researcher, Agency for Defense Development (ADD), Yuseong, P.O.Box 35, Daejeon, 305-600, Korea/mountrees@gmail.com.
[4] Graduate Student, Department of Aerospace Engineering, Korea Advanced Institute of Science and Technology (KAIST), Kuseong-dong, Yuseong-gu, Daejeon, 305-701, Korea/mgseo@fdcl.kaist.ac.kr.
[5] Professor, Department of Aerospace Engineering, Korea Advanced Institute of Science and Technology (KAIST), Kuseong-dong, Yuseong-gu, Daejeon, 305-701, Korea/mjtahk@fdcl.kaist.ac.kr.

# I. Introduction

In recent missile autopilot design, the consideration of nonlinear control methodologies has grown in order to take into account the model nonlinearities and the variations of model parameters according to the changes of the operating points. Above all, the feedback linearization methodology [1] has been widely used to address the flight control problems, the missile control systems [2-16], and the aircraft control systems [18-23] because of its many benefits, such as simple application to a class of nonlinear systems and a straightforward means of controller design for nonlinear systems by simply canceling the nonlinearities and imposing the desired linear dynamics. However, in the application of feedback linearization methodology to the design of the missile acceleration controller, there exist some challenging problems: the handling of nonminimum phase phenomena and the inaccuracy of the model due to the aerodynamic coefficient uncertainties. In order to overcome these difficulties, there have been a number of approaches over the past decades.

As a method for handling the nonminimum phase phenomena, the two-time scale separation method was proposed in [2-3]. The authors in [4-5] suggested the concept of output redefinition in order to obtain a minimum phase system. In [6-9], a cascade form of controller design methodology based on feedback linearization and classical control theory was studied. In this approach, the angle-of-attack control, which is known to be the minimum phase, was performed based on feedback linearization, and then, a classical controller was added around it in order to control the acceleration. In [10-12], the approximate system, which neglects the terms causing the nonminimum phase characteristics, was used to design the acceleration controller. In [13], the missile acceleration autopilot was designed using the input-output pseudo-feedback linearization with an augmented lateral acceleration signal to relieve the effect of the nonminimum phase. The eigenstructure assignment [15-16] and the asymptotic output tracking [11, 17] approaches were also applied to the nonminimum phase problem.

In order to overcome the difficulty related to the model uncertainties, feedback linearization combining an adaptive scheme based on neural networks was proposed in [14]. In [19-20], an outer loop controller based on quantitative feedback theory was considered in order to improve the robust performance. $\mu$ synthesis was also used in conjunction with the feedback linearization control in [22-23].

This paper suggests a different approach to tackling the described problems: using a cascade control structure based on the approximate feedback linearization methodology with a time-delay adaptation scheme. Ignoring the nonminimum characteristic terms in the original system introduces an approximate system [10-12, 18]. This approximate system is known to be the minimum phase, and the output of the achieved approximate system can be regarded as the acceleration signal caused by the angle-of-attack. Then, the input-output feedback linearization is applied to this approximate system for an inner loop controller, and a classical controller, such as the PI (proportional-integral) controller, is augmented for an outer loop in order to compensate for the acceleration tracking error due to ignoring terms in the approximate system as compared with the original system. According to the previous studies [10, 12], the performance of the controller based on the approximate feedback linearization may be severely degraded in the presence of the model uncertainty because of its controller structure, that is, it requires the second derivative of uncertain aerodynamic parameters with respect to state variables and such a derivative process may cause excessive model uncertainties. Hence, it is necessary to augment an adaptive law so as to improve the approximate feedback linearization to compensate for the model uncertainties. In this study, the time-delay approximation technique, which is the key idea of the time-delay control methodology [24], is used for deriving the proposed adaptation scheme, which is called the time-delay adaptive law. From previous studies [25-26], it has been indicated that the time-delay approximation technique is an efficient and a practical estimation means for the model uncertainties. In order to examine the performance of the proposed method, a number of nonlinear simulations involving imposing a step command are performed. Furthermore, the proposed method is tested with a target intercept scenario in a 6-DOF nonlinear simulation.

This paper starts with a description of the considering missile model in Section II. Section III discusses the stability analysis of the zero dynamics for the acceleration control and the derivation of the approximate system used. The proposed autopilot design methodology is proposed in Section IV. In Section V, simulation studies are conducted in order to examine the performance of the proposed method. Finally, we conclude our investigation in Section VI.

## II. Missile Model

This section discusses a nonlinear missile model for an acceleration controller design. In skid-to-turn (STT) cruciform-type missile systems, the missile motion can be decoupled into two perpendicular channels: the pitch and the yaw motion. Therefore, a nonlinear missile model in the pitch motion is considered as shown in Fig. 1. The equation of motion is expressed as

$$\dot{\alpha} = \frac{QS}{mV}\left[C_{Z_0}(M,\alpha) + C_{Z_\delta}(M,\alpha)\delta + \Delta C_Z\right] + q$$

$$\dot{q} = \frac{QSl}{I_{yy}}\left[C_{M_0}(M,\alpha) + C_{M_q}(M)\frac{ql}{2V} + C_{M_\delta}(M,\alpha)\delta + \Delta C_M\right] \quad (1)$$

$$a_Z = \frac{1}{m}QS\left[C_{Z_0}(M,\alpha) + C_{Z_\delta}(M,\alpha)\delta + \Delta C_Z\right]$$

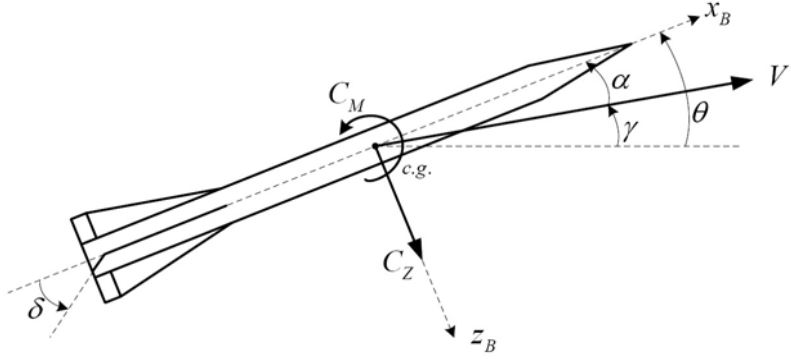

Fig. 1 The missile axes and the definition of dynamic variables.

where $\alpha$ and $q$ are the angle-of-attack and the body pitch rate, which are the state variables. The control fin, denoted by $\delta$, is the control input. The acceleration component of $Z$ (i.e., $a_Z$) is to be controlled. The notations of $M$, $Q$, and $V$ represent Mach number, dynamic pressure, and velocity, respectively. The other parameters are the specifications of the missile: reference area $S$, reference length $l$, mass $m$, and pitching moment of inertia $I_{yy}$.

The aerodynamic coefficients described by $C_{Z_0}$, $C_{Z_\delta}$, $C_{M_0}$, $C_{M_q}$, and $C_{M_\delta}$ are dependent on Mach number and angle-of-attack, and these parameters can be mostly regarded as continuous functions of their arguments. The aerodynamic coefficients are measured from wind tunnel tests, and these values may contain some errors as compared with true values because of the imperfection of the measurements. These errors can be modeled as multiplicative uncertainties:

$$C_{(\cdot)}^{pert} = (1 + \Delta_{pert})C_{(\cdot)} \quad (2)$$

where $C_{(\cdot)}$ are the true aerodynamic coefficients, and $C_{(\cdot)}^{Pert}$ are the measured aerodynamic coefficients. $\Delta_{pert}$ represents the admissible uncertainties, and their maximum values are about 0.2 ~ 0.3. The coefficients $\Delta C_Z$ and $\Delta C_M$ represent the aerodynamic coefficients due to the cross-coupling effect of the missile motion, which are regarded as additive uncertainties and are mostly continuous functions of Mach number and angle-of-attack.

## III. Derivation of Approximate System

Applying the feedback linearization methodology to the missile acceleration controller design is generally limited because the system in Eq. (1) is a nonminimum phase and the controller based on the feedback linearization usually requires precise knowledge of the system model. Before the autopilot design step, in order to relieve these difficulties, this section derives an approximate system [10, 12] to be the minimum phase. The handling of the model uncertainties will be discussed in the Section IV.

### A. Approximate System

Since a straightforward application of the feedback linearization methodology to the system for the acceleration control results in unstable zero dynamics, we introduce an approximate system [10, 12] to be a minimum phase, eliminating the $C_{Z_\delta}\delta$ term in $\dot{\alpha}$, as shown in Eq. (1), due to $\left|C_{Z_\delta}(M,\alpha)\delta\right| \ll \left|C_{M_\delta}(M,\alpha)\delta\right|$ and introducing an approximate output denoted by $\bar{a}_Z$, which is the acceleration signal caused by the angle-of-attack.

$$\dot{\alpha} = \frac{QS}{mV}C_{Z_0}(M,\alpha) + q$$
$$\dot{q} = \frac{QSl}{I_{yy}}\left[C_{M_0}(M,\alpha) + C_{M_q}(M)\frac{ql}{2V} + C_{M_\delta}(M,\alpha)\delta\right] \quad (3)$$

$$\bar{a}_Z = \frac{1}{m}QSC_{Z_0}(M,\alpha) \quad (4)$$

In this system, the relative degree is two, and this is the same as its system order. Hence, there are no internal dynamics, and feedback linearization can be easily achieved. In this study, the controller will be designed based on the approximate system, and it will be applied to the original system.

## IV. Autopilot Design

This section discusses the autopilot design for the acceleration control using a cascade control structure [6-9]: the inner loop controller and the outer loop controller. The inner loop controller is formulated based on the approximate system, using the feedback linearization methodology, and it controls the approximate output (i.e., the acceleration signal caused by the angle-of-attack). The time-delay adaptive law is used to compensate for the model uncertainties. A classical PI controller is augmented in the outer loop to control the original output (i.e., total acceleration).

### A. Inner Loop Controller Design

For notational convenience, we introduce new variables as follows:

$$x_1 = \alpha, \qquad x_2 = q, \qquad u = \delta, \qquad y = \bar{a}_Z \tag{5}$$

where $x_1$ and $x_2$ represent the state variables. The control input and the output of the approximate system are denoted by $u$ and $y$, respectively. Using these variables, the approximate system as shown in Eqs. (3) and (4) can be rewritten in parts of the state variables and the control input, separately.

$$\begin{aligned}\dot{x}_1 &= f_1(x_1) + x_2 \\ \dot{x}_2 &= f_2(x_1, x_2) + g_2(x_1)u \\ y &= h(x_1)\end{aligned} \tag{6}$$

where,

$$\begin{aligned}f_1(x_1) &\triangleq \frac{QS}{mV} C_{Z_0}(M, \alpha) \\ f_2(x_1, x_2) &\triangleq \frac{QSl}{I_{yy}} \left[ C_{M_0}(M, \alpha) + C_{M_q}(M) \frac{ql}{2V} \right] \\ g_2(x_1) &\triangleq \frac{QSl}{I_{yy}} C_{M_\delta}(M, \alpha) \\ h(x_1) &\triangleq \frac{QS}{m} C_{Z_0}(M, \alpha)\end{aligned} \tag{7}$$

Based on the input-output feedback linearization, in order to find an explicit relationship between the output of the approximate system $y$ and the control input $\delta$ from Eq. (6), we differentiate the output of the approximate system until the control input appears.

$$\dot{y} = \frac{\partial h(x_1)}{\partial x_1} \left[ f_1(x_1) + x_2 \right] \tag{8}$$

$$\ddot{y} = \frac{\partial^2 h(x_1)}{\partial x_1^2} \left[ f_1(x_1) + x_2 \right]^2 + \frac{\partial h(x_1)}{\partial x_1} \left\{ \frac{\partial f_1(x_1)}{\partial x_1} \left[ f_1(x_1) + x_2 \right] + f_2(x_1, x_2) + g_2(x_1)u \right\} \tag{9}$$

After taking the time-derivative of $y$ twice (i.e., the relative degree is two), the control input directly appears in Eq. (9). For notational convenience, this equation can be rewritten using shorthand notation as follows:

$$\ddot{y} = f_3 + g_3 u \tag{10}$$

where

$$f_3 \triangleq \frac{\partial^2 h(x_1)}{\partial x_1^2}\left[f_1(x_1)+x_2\right]^2 + \frac{\partial h(x_1)}{\partial x_1}\left\{\frac{\partial f_1(x_1)}{\partial x_1}\left[f_1(x_1)+x_2\right]+f_2(x_1,x_2)\right\}$$
$$g_3 \triangleq \frac{\partial h(x_1)}{\partial x_1}g_2(x_1) \tag{11}$$

This autopilot design method requires the second derivatives of the aerodynamic coefficients with respect to the angle-of-attack in Eq. (10). Since the aerodynamic coefficients mostly contain uncertainties such as the aerodynamic coefficient perturbations due to the imperfection of wind tunnel measurements and the cross-coupling effect of the missile model, the second derivatives of uncertain aerodynamic coefficients may introduce large model uncertainties [10, 12]. Therefore, these model uncertainties should be carefully considered in the autopilot design step. Accordingly, from Eq. (10), taking the model uncertainties into consideration, the dynamic equation can be expressed as:

$$\dddot{y} = f_3 + g_3 u + \Delta \tag{12}$$

where $\Delta$ represents all possible model uncertainties. In order to design the control law, let the desired tracking error dynamics be first defined as follows:

$$\ddot{e} + K_1 \dot{e} + K_2 e = 0 \tag{13}$$

where $e = \bar{a}_{Z,c} - \bar{a}_Z$. The notations of $K_1$ and $K_2$ denote the controller gains. Based on the feedback linearization methodology, substituting Eq. (12) into Eq. (13), we have the control law as follows:

$$u = g_3^{-1}\left(-f_3 + \dddot{\bar{a}}_{Z,c} + K_1 \dot{e} + K_2 e - \Delta\right) \tag{14}$$

In this equation, the control law contains the unknown term (i.e., the model uncertainties) described by $\Delta$. If this term is neglected in the control law, the degradation of the tracking performance is predicted [10, 12]. Therefore, the model uncertainties should be adapted to improve the tracking performance. The proposed adaptive law will be discussed the following subsection.

**B. Time-delay Adaptive Law**

This section discusses the proposed adaption scheme, which is called the time-delay adaptive law. This is a practical and an efficient method to estimate model uncertainties and it is formulated based on the time-delay approximation technique [24].

If a time-varying function $f(t)$ is continuous for $0 \leq t \leq b$, the following approximation is possible for a sufficiently small time-delay $L$:

$$f(t) \simeq f(t-L) \tag{15}$$

This is the core idea of the time-delay approximation. Using this scheme, the model uncertainties can be estimated as:

$$\hat{\Delta}(t) = \Delta(t-L) \tag{16}$$

Substituting Eq. (12) into Eq. (16), the time-delay adaptive law is formulated as follows:

$$\hat{\Delta}(t) = \ddot{y}(t-L) - f_3(t-L) - g_3(t-L)u(t-L) \tag{17}$$

This adaptive law requires the time-delayed model, state variables, and control input information, which are obtained using the single-lag system:

$$f(t-L) \triangleq \frac{1}{\tau_d s + 1} f(t) \tag{18}$$

where $\tau_d$ is the time constant related with the delayed signal and is a design parameter. Employing the time-delay adaptive law, the adaptation error decreases as the time constant of the single-lag system decreases. In addition, since the single-lag system can act as a first-order low-pass filter, the high frequency signal in the delayed information can be rejected.

**C. Outer Loop Controller Design**

Applying the control law, as designed in the previous subsection and, formulated based on the approximate system, to the original system, a tracking error in the acceleration is expected due to ignoring certain terms in deriving the approximate system. In other words, the inner loop controller provides the desired acceleration signal caused by the angle-of-attack, not the total acceleration. In order to compensate for this tracking error, a cascade control structure [6-9] is used, as shown in Fig. 2. The outer loop controller based on a classical PI control methodology is added to the inner loop controller.

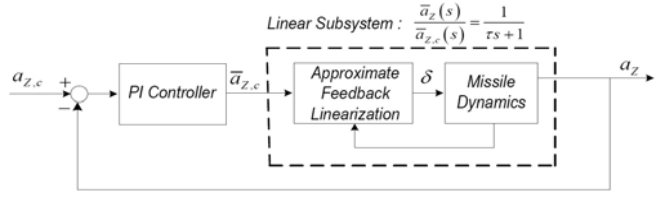

Fig. 2 The cascade control structure for the acceleration control.

We consider the following PI controller for the outer loop to compensate for the tracking error.

$$G(s) = K_P + \frac{1}{s}K_I \tag{19}$$

where $K_P$ and $K_I$ denote the proportional and the integral gains, respectively. From Fig. 2, the inner loop controller for tracking the approximate acceleration can be regarded as the linear subsystem, and its transfer function can be approximated as follows:

$$\frac{\bar{a}_Z(s)}{\bar{a}_{Z,c}(s)} = \frac{1}{\tau s + 1} \tag{20}$$

where $\bar{a}_{Z,c}$ is the command, and $\tau$ is the time constant of the inner loop controller. Since the acceleration is mostly proportional to the approximate acceleration, it can be also approximated as:

$$a_Z(s) = K\bar{a}_Z(s) \tag{21}$$

From Fig. 2, the transfer function of the acceleration command to the acceleration response can be determined as the following by using Eqs. (19), (20), and (21).

$$\frac{a_Z(s)}{a_{Z,c}(s)} = \frac{\frac{K}{\tau}(K_P s + K_I)}{s^2 + \frac{(KK_P + 1)}{\tau}s + \frac{KK_I}{\tau}} \tag{22}$$

Let the reference closed-loop characteristics be defined as follows:

$$G_R(s) = \frac{\omega_R^2}{s^2 + 2\zeta_R \omega_R s + \omega_R^2} \tag{23}$$

where $\omega_R$ and $\zeta_R$ represent the reference natural frequency and damping. Then, the controller gains for the outer loop are determined for Eq. (22), which follows the above-referenced closed-loop characteristics as follows:

$$K_P = \frac{2\zeta_R \omega_R \tau - 1}{K}, \qquad K_I = \frac{\tau \omega_R^2}{K} \tag{24}$$

## V. Simulation Results

This section investigates the performance of the proposed control methodology through a number of numerical simulations. First, the inner loop controller, based on the approximate system and the feedback linearization methodology with the time-delay adaptive scheme, is tested by imposing a step command. Then, the performance of the overall controller (inner loop controller with outer loop PI controller) for the acceleration control is investigated. The relative stability of the proposed controller is determined using the numerical method. Finally, the proposed controller is tested by a target intercept scenario in a 6-DOF simulation.

In all simulations, the second-order linear command shaping filter is used in order for obtaining differential commands, and the second-order actuator model is considered; the natural frequency $\omega_a = 180 \, \text{rad/s}$, the damping ratio $\zeta_a = 0.7$, the control fin limit $\delta_{\text{lim}} = \pm 30°$, and the control fin rate limit $\dot{\delta}_{\text{lim}} = \pm 450°/\text{sec}$. The parameters of the proposed controller are designed as follows: $K_1 = 30$ and $K_2 = 15^2$ for the inner loop controller, $K_P = 0.61$ and $K_I = 13.2$ for the outer loop controller, and $\tau_d = 0.02$ for the time-delay adaptive law.

**A. Approximate Acceleration Control (Inner Loop Only)**

At Mach 2.5, the inner loop controller (without the outer loop) is tested by imposing a step command for the nominal case and the consideration of the model uncertainties case. Fig 3 presents the results of the approximation acceleration control for the nominal case. It shows good tracking of the desired approximate acceleration. Fig 4 shows the results in the presence of the model uncertainties, such as 30% multiplicative uncertainties ($\Delta_{Pert} = 0.3$) in aerodynamic coefficients from wind-tunnel measurement error and the additive aerodynamic uncertainties due to the cross-coupling effect of the missile motion. Without the adaptation (FL, feedback linearization), there exists a bias tracking error, and with the adaptation (TDAFL, time-delay adaptive feedback linearization), the controller provides sound tracking performance, even in the presence of the model uncertainties. Fig 4c) presents the estimate of the model uncertainties using the time-delay adaptive law. It shows the sound estimation performance of the proposed adaptive law.

**B. Acceleration Control (Inner Loop with Outer Loop)**

The performance of the overall controller for the acceleration control is tested under the nominal case and with the model uncertainties. A step command and the model uncertainties identical to those in the previous section are considered in this simulation. Figs. 5 and 6 show the results of the acceleration control for the nominal case and the model uncertainties case, respectively. In these figures, the dashed line is the response of the approximate acceleration, and the solid line represents the response of the acceleration. Both results indicate that the proposed controller can provide good tracking of the desired acceleration, even if large model uncertainties exist.

## VI. Conclusion

In this investigation, a cascade form of the missile acceleration controller was proposed. The inner loop controller was designed by applying feedback linearization methodology to the approximate system, which was derived by ignoring the nonminimum phase characteristics of the original system, and the time-delay adaptive law was augmented in order to reject the model uncertainties. Then, a classical outer loop PI controller was added to provide the desired acceleration command. The proposed method was tested by a number of nonlinear simulations, and the parameters showing the relative stability of the controller, such as the gain margin and the phase margin, were determined using the numerical calculations. The results indicated that the proposed method provided good tracking performance, even in the presence of model uncertainties and model nonlinearities, and might be suitable for the application of nonminimum phase systems. Employing the time-delay adaptive law, the model uncertainties were estimated quite well. Also, the proposed controller was tested with a target intercept scenario in a 6-DOF nonlinear simulation. It showed that the proposed method can be applied to the challenging issues of the missile acceleration controller.